\documentclass[journal=jacsat,manuscript=article]{achemso}

\usepackage[version=3]{mhchem} 
\usepackage[T1]{fontenc}       
\usepackage{array}
\usepackage{booktabs}
\usepackage{listings}
\usepackage{lmodern}
\usepackage{mathpazo}
\usepackage{microtype}
\usepackage{physics}
\usepackage{bm}

\usepackage{graphicx}
\usepackage{float}

\usepackage{amsthm}
\usepackage{amsmath}
\usepackage{amsfonts}

\usepackage{commath}

\author{Yuan Yang}
\affiliation{Department of Chemistry and Chemical Biology, Harvard University, Cambridge, MA, 02138, United States}
\author{Grigory Kolesov}
\affiliation{John Paulson School of Engineering and Applied Sciences, Harvard University, Cambridge, MA, 02138, United States}
\author{Lucas Kocia}
\affiliation{Department of Physics, Tufts University, Medford, MA, 02155, United States}
\author{Eric J. Heller}
\email{heller@physics.harvard.edu}
\affiliation{Department of Chemistry and Chemical Biology, Harvard University, Cambridge, MA, 02138, United States}
\alsoaffiliation{Department of Physics, Harvard University, Cambridge, MA, 02138, United States}

\title[An \textsf{achemso} demo]
  { Reassessing Graphene Absorption and Emission Spectroscopy}

\abbreviations{Graphene, Optical Absorption, Optical Emission}

\begin{document}







\begin{abstract}
We present a new paradigm for understanding optical absorption   and hot electron dynamics experiments in graphene.  Our analysis pivots on assigning proper importance   to phonon assisted indirect processes and bleaching of direct processes. We show indirect processes figure in  the excess absorption in the UV region. Experiments which were  thought to indicate ultrafast relaxation of electrons and holes, reaching a thermal distribution from an extremely non-thermal one in under $5-10$ fs,  instead  are explained by the nascent  electron and hole distributions produced   by   indirect transitions.  These need no relaxation or ad-hoc energy removal to agree  with the observed emission spectra and fast pulsed absorption spectra. The fast emission following pulsed absorption  is  dominated by phonon assisted  processes, which vastly outnumber direct ones and are always available, connecting any electron with any hole any time.   Calculations are given, including explicitly calculating the magnitude of indirect processes, supporting these views.
\end{abstract}

\begin{figure}[H]
\includegraphics[width=11cm]{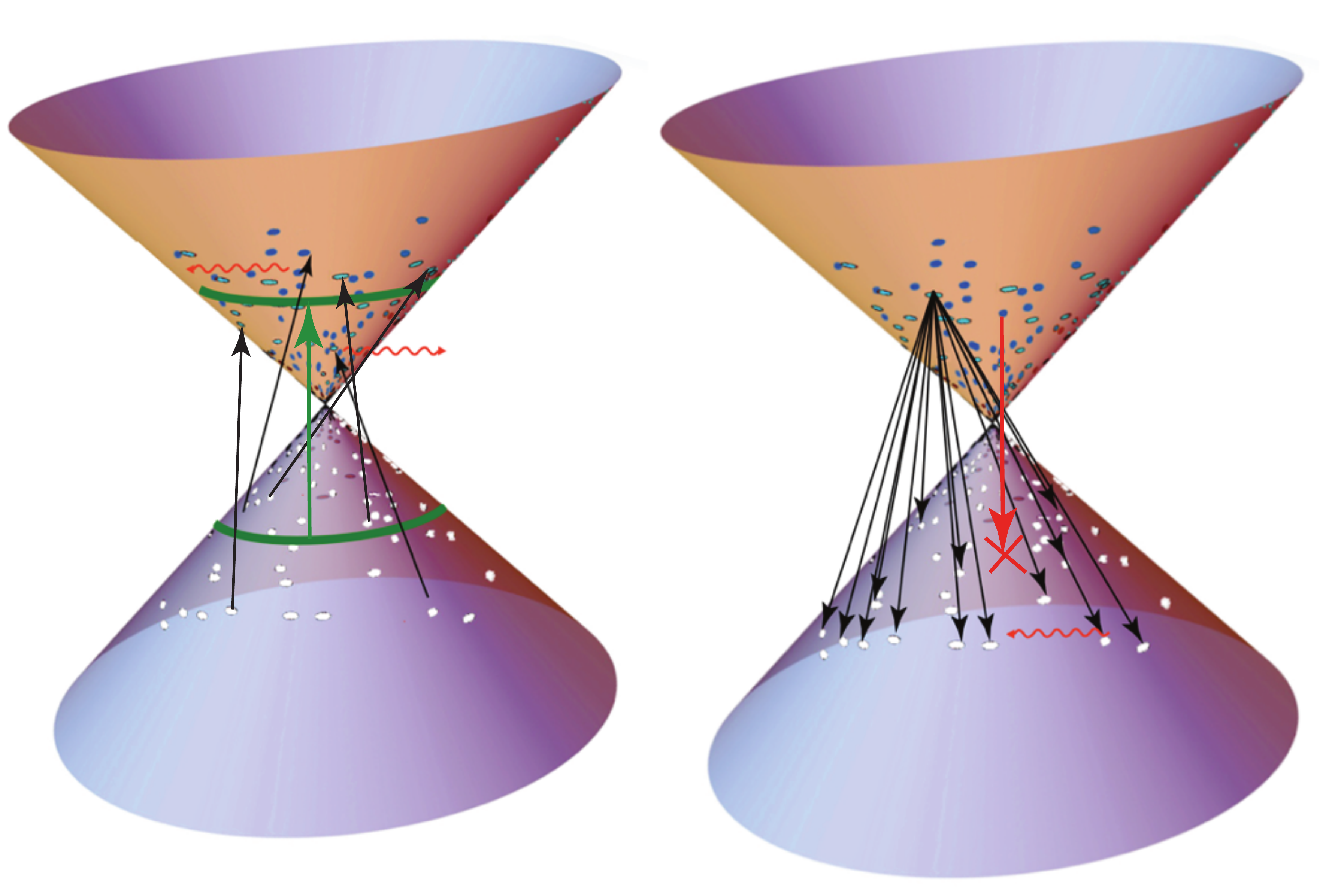}
\caption{An illustration of mostly indirect optical absorption and emission processes in graphene: The left image shows two  absorption mechanisms. The green arrow linking two green circles is a direct phononless absorption process from $-\hbar \omega/2$ in the valence band  to $\hbar \omega/2$ in the conduction band, and the black arrows linking holes and electrons  are phonon assisted indirect absorption processes at the same vertical $\hbar \omega$. The right image shows the emission mechanism. Any electron in the conduction band is ready for emission to any hole in the valence band   by a phonon   assisted transition. It does not need to await perfect momentum alignment of randomly distributed electrons and holes. This is not the reverse of absorption, where a vacant conduction band state awaits every valence state.    Conduction electrons are almost always Pauli blocked for direct  emission (red arrow with X).}
\label{absorption_emission_barePlus}
\end{figure}
\section{Introduction}
Graphene, with its $sp^2$-hybridized honeycomb two-dimensional carbon lattice consisting of conjugated hexagonal cells, shows extraordinary optical properties because of its dimensionality and unique electronic band structure\cite{Novoselov666}. As an atomically thin two-dimensional carbon material, graphene is  used for transparent electrodes and optical display materials.  It has also been applied in opto-electronics such as photodetectors, optical modulators, and so on\cite{Bonaccorso2010,kim2009large,wang2008transparent,bae2010roll,vicarelli2012graphene,ju2011graphene,liu2011graphene,ren2012terahertz,lee2012switching,furchi2012microcavity,zhang2015ultrasensitive}. A proper understanding of the carrier dynamics in graphene is key to its potential applications in high-speed photonics and optoelectronics. Many theoretical works concern the electron dynamics in graphene\cite{Bonaccorso2010,Mak20121341,Heller:2015aa}. 


In condensed matter theory, the electronic transition moments connecting valence and conduction band states are traditionally taken to be constant, independent of phonon displacement.  However  we have found most of graphene spectroscopy falls into place only when transition moments are freed to depend on phonon coordinates, as they must do.  (For example, the polarizablity of a material would remain constant under arbitrary lattice distortion if the electronic transition moments remain unchanged under those distortions). Recently, it was shown that indirect transitions  induced by coordinate dependence of the transition moment  are responsible for the great intensity of some Raman overtones \cite{heller2016theory}.  This puts Raman scattering in graphene in line with traditional  2nd order Kramers-Heisenberg-Dirac Raman scattering,  in use since 1925-27.  In the present paper, coordinate dependence of the transition moment and indirect processes   again  play a central role for femtosecond pulse-probe absorption and femtosecond  pulsed emission experiments, as well as traditional absorption spectroscopy. We show that no ultrafast relaxation is necessary or implied by a variety of experiments.
  
A significant body of work claims or assumes that ultrafast (5-10 fs) electron-electron relaxation follows ultrashort pulsed excitation.   This seemed to be an obvious inference: in a variety of experiments, one looked within femtoseconds after (one supposed) creating  an extremely non-equilibrium initial population, only to find the electrons behaving relaxed and even thermalized.   For example, upon short pulsed excitation, graphene samples produce fast,  readily observable light emission, appearing to be coming from a relaxed or even thermal  distribution as soon as it can be seen,  on the 7-20 fs time scale\cite{PhysRevB.82.081408,Mak20121341,PhysRevLett.101.196405}. Ultrafast pump-probe absorption experiments similarly see  spectra that appear to have no resemblance to the assumed narrow ranges of populated electrons and holes\cite{PhysRevLett.102.086809}.

Nonetheless there have long been clouds on the horizon of the ultrafast landscape.  No experiment has ever caught a system in the act of the supposed ultrafast relaxation, nor any vestige of the putative ultrafast component.   The  fastest relaxation times directly measured  in  experiments have been in the 100-300 fs range, often attributed in the ultrafast literature to fast, but not ultrafast electron-phonon inelastic events.  However  saturable absorption (SA) experiments reveal unambiguously that the \textit{fastest} timescale for electronic relaxation is 100-200 fs, which agrees with earlier  femtosecond pump-probe measurement on pyrolytic graphite\cite{moos2001anisotropy,kumar2009femtosecond,bao2011monolayer}.   The timescale for relaxation from an extremely nonequilibrium, saturated   narrow band of energies to full thermalization could  not be 10-20 fs or shorter, if saturation recovery takes 100-300 fs.  

Another cloud threatening  the ultrafast  narrative could be called the missing energy conundrum.  In Lui et. al. \cite{PhysRevLett.105.127404}, it is mentioned that the vertical energy per conduction electron at 1.5 eV (0.75 eV for the conduction band electrons)  corresponds to an electron temperature of 9000 Kelvin.  Once the assumption is made that the electrons and their emission are thermal just a few femtoseconds after the pulse, one is forced to  arbitrarily remove 2/3 of the  energy that has just been supplied to the electrons and use  3000 Kelvin electron temperature instead of 9000 to fit the emission. This is a serious defect, since the fastest process and the only one effective on the few femtosecond timescale,  namely e-e scattering,  cannot change the average energy per electron.  The temperature becomes an adjustable fitting parameter, even though it should have been non-adjustable.  Carrier multiplication might be suspected, and  was indirectly inferred  and modeled as the only plausible explanation for what seemed to be 10 fs relaxation\cite{brida2013ultrafast}.  However when carrier multiplication  was was actually measured, it wasn't found to be present\cite{gierz2013snapshots}. 

Seemingly in favor of the ultrafast relaxation idea are the beautiful experiments measuring electron coherence as seen by D - mode Raman scattering from a localized source, made visible after elastic backscattering from edges\cite{brida2013ultrafast}. If the source of the conduction band electron was more than 8 nm   round trip from the edge, or about 8 fs, the D mode lost intensity due to lack of coherence with the hole.  However, this was properly viewed as a coherence length and dephasing issue, and not a measurement of the complete electronic relaxation time by any means.

Experimental results of disparate types  fall into one unified picture if the "thermal" electron and hole distribution is produced not by any relaxation, but nascently at $t=0$ by a dominance of indirect transitions which, although  present even for weak radiation, take over from the easily bleached direct transitions in a bright pulse\cite{bao2009atomic}.   To state it plainly, the electron-hole distribution is born pre-"relaxed" in bright pulsed absorption.  This new narrative  is competed by careful consideration of how   emission  takes place, which  is  also  by phonon assisted channels that vastly outnumber elastic  processes (figure~\ref{absorption_emission_barePlus}).   Though individually weaker than the elastic channels,  the inelastic pathways are  available from any electron to any hole at any moment.  Direct emission on the other hand requires waiting for  perfect momentum coincidence of randomly distributed electrons and holes. See figure~\ref{absorption_emission_barePlus}. Following the lead of  indirect absorption plus indirect emission  gives excellent agreement with experiments, without the need for arbitrary excited state energy removal, or  any excited state relaxation at all on the femtosecond timescale.  Although this is a long way from the prevailing consensus, it  agrees with   direct measurements of electron-electron relaxation rates from saturation experiments, mentioned   above, and is free from the clouds and conundrums that the prevailing views are laboring under.

  
 Regarding ordinary CW light absorption, excess absorption over the ``universal'' value develops in the UV region\cite{Nair1308,PhysRevLett.101.196405,Wang206,Li2008}.   We show here by explicit calculations that phonon-assisted transitions play an increasingly important role as the laser frequency enters the UV region, contributing to the excess absorption in the UV.

In what follows,   by new results for  graphene UV absorption are  discussed first, where phonon assisted processes are key. Then, we address the subject of fast spontaneous emission following bright pulsed excitation, showing that phonon assisted processes explain the spectra without any relaxation. Finally we show   that   seemingly ultra-rapid relaxation  seen in pump-probe absorption experiments are instead the result of nascently produced electron and hole distributions via phonon assisted processes.


\section{Absorption Spectrum}
\subsection{Theoretical Background}
In the Supplemental Material, an expression is derived for the  absorption cross section from an initial state $\ket{i}$ to a final state $\ket{n}$ as 

\begin{equation}
\frac{\textrm{(Energy/unit time)absorbed by the lattice(i$\rightarrow$n)}}{\textrm{Energy flux of the radiation field}}
\end{equation}

 The absorption cross section from an initial state $\ket{i}$ to a final state $\ket{n}$ can be written in terms of the transition moment as
\begin{equation}
\label{cross_section_i_n}
\sigma_{i,n}^{\hat{\boldsymbol{\epsilon}}}=\frac{4\pi^2\hbar}{m_e^2\omega}\frac{e^2}{\hbar c}\mid\bra{\chi_{\boldsymbol{m}_n}(\boldsymbol{\xi})}\mu_{\boldsymbol{q}_c,\boldsymbol{q}_v}^{\hat{\boldsymbol{\epsilon}}}(\boldsymbol{\xi})\ket{\chi_{\boldsymbol{m}_i}(\boldsymbol{\xi})}\mid^2\delta(E_n-E_i-\hbar\omega)
\end{equation}
with the phonon coordinate dependent transition moment
$
\mu_{\boldsymbol{q}_c,\boldsymbol{q}_v}^{\hat{\boldsymbol{\epsilon}}}(\boldsymbol{\xi})=\bra{\phi_{\boldsymbol{q}_v\boldsymbol{q}_c}(\boldsymbol{\xi};\boldsymbol{r})}\boldsymbol{D}^{\hat{\boldsymbol{\epsilon}}}\ket{\phi(\boldsymbol{\xi};\boldsymbol{r})}_{\boldsymbol{r}},
$
 connecting the Born-Oppenheimer valence and conduction band electronic states, some with different phonon occupations, given by an matrix element of the dipole operator  $\boldsymbol{D}^{\hat{\boldsymbol{\epsilon}}}$  over the electronic states at the given nuclear positions.  The wavefunction $\ket{\chi_{\boldsymbol{m}_i}(\boldsymbol{\xi})}$ is a particular nuclear wavefunction with phonons labeled by  $\boldsymbol{m}_i$.

\subsection{Computational Methods}

We use SIESTA to perform DFT calculations to get vertical phononless absorption and phonon-assisted absorption. To get meaningful physical quantities, we need wavefunctions having complete periods in our finite lattice, an 80 by 80 graphene supercell. $\Gamma$ point wavefunctions in the supercell to calculate both absorptions.

For direct, vertical absorption, we consider all electronic transitions from valence band to conduction band at the $\Gamma$ point of the supercell Brillouin zone, and with electronic wavefunctions we can compute $\mu_{\boldsymbol{q}_c,\boldsymbol{q}_v}^{\hat{\boldsymbol{\epsilon}}}(\boldsymbol{\xi}_0)$.
For  indirect phonon-assisted absorption, we  calculate $\frac{\partial{\mu_{\boldsymbol{q}_c,\boldsymbol{q}_v}^{\hat{\boldsymbol{\epsilon}}}(\boldsymbol{\xi})}}{\partial \xi_j}\mid_{\boldsymbol{\xi}=\boldsymbol{\xi}_0}$, i.e. the change of transition moment under lattice distortion, by a finite difference method.

\begin{equation}
\label{TM_Derivative_Finite_Difference}
\frac{\partial{\mu_{\boldsymbol{q}_c,\boldsymbol{q}_v}^{\hat{\boldsymbol{\epsilon}}}(\boldsymbol{\xi})}}{\partial \xi_j}\mid_{\boldsymbol{\xi}=\boldsymbol{\xi}_0}=\frac{{\mu_{\boldsymbol{q}_c,\boldsymbol{q}_v}^{\hat{\boldsymbol{\epsilon}}}(\boldsymbol{\xi+\delta\xi})}-\mu_{\boldsymbol{q}_c,\boldsymbol{q}_v}^{\hat{\boldsymbol{\epsilon}}}(\boldsymbol{\xi-\delta\xi})}{2\mid\boldsymbol{\delta\xi}\mid}
\end{equation}
The numerator in Eq (\ref{TM_Derivative_Finite_Difference}) reads as the change of transition moment from an electronic state $\boldsymbol{q}_v$ to $\boldsymbol{q}_c$ when lattice is distorted from $-\boldsymbol{\delta\xi}$ configuration by $+2\boldsymbol{\delta\xi}$ to $\boldsymbol{\delta\xi}$.  We use  a diabatic approximation to the electronic states when the lattice is distorted by $+2\boldsymbol{\delta\xi}$ from $-\boldsymbol{\delta\xi}$, connecting the maximally overlapping adiabatic states, which have been re-computed after the change in nuclear displacements. (Recall that some lattice symmetry has been broken to make the displacements, so the old set of good quantum numbers do not otherwise make the connection obvious). 



\subsection{Absorption in the Near-IR-to-UV Spectral Region}

\begin{figure}[H]
\includegraphics[width=13cm]{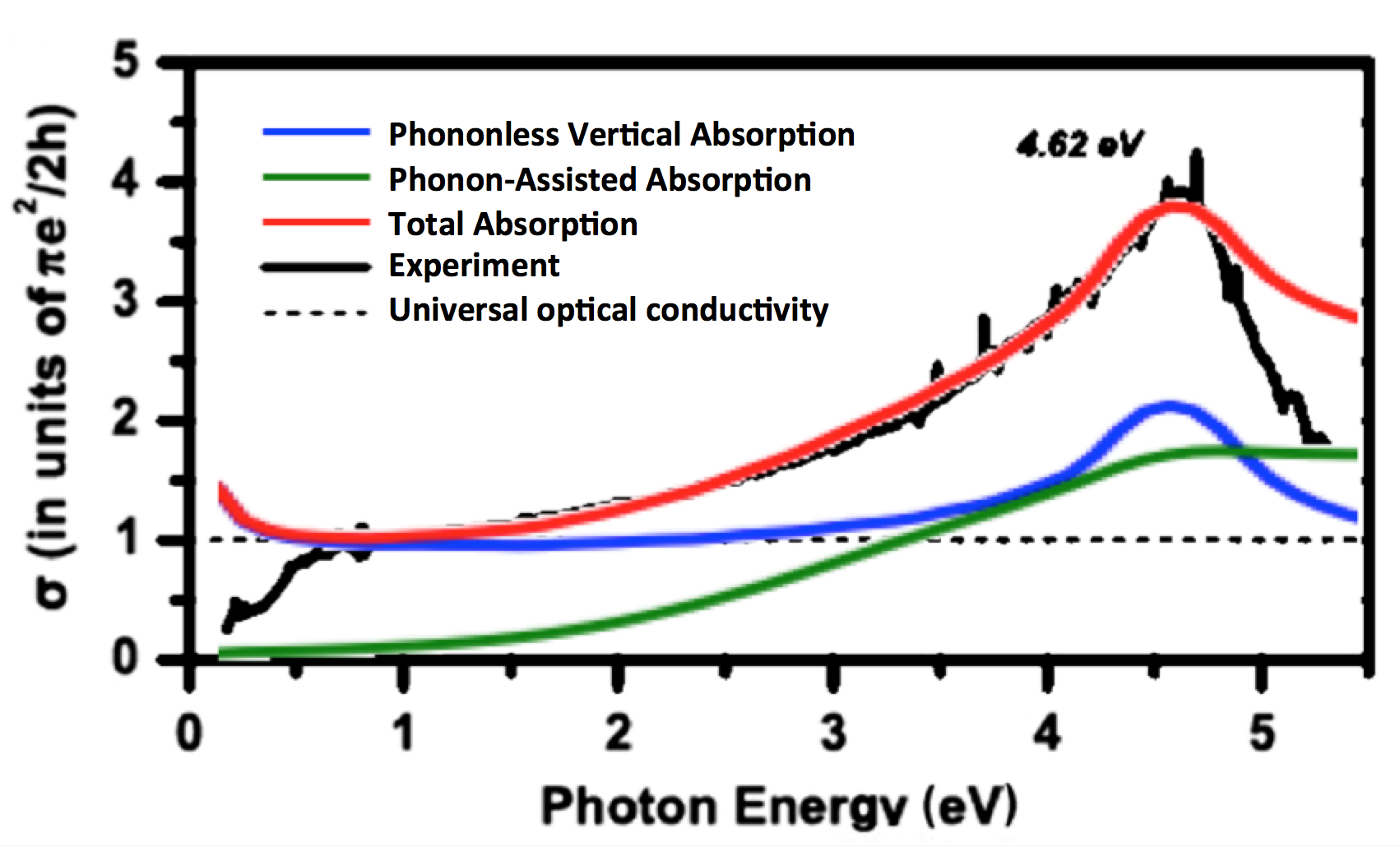}
\caption{Calculated absorption curve vs experiment: The dark solid line is the experimental absorption curve adapted from the work of Mak etc.\cite{PhysRevLett.106.046401}. The dash line is the universal optical conductivity. The blue line is the calculated absorption contributed by direct  phononless transitions, the green line is the calculated absorption from indirect phonon-assisted processes, and the red line is the total absorption by summing the phononless transitions and phonon-assisted transitions. The coordinate dependence of the transition moment enables the indirect processes,}
\label{absorption_fit}
\end{figure}

Graphene displays universal absorption in the near-IR   region of $0.5-1.5 eV$, and a slow, at first  quadratic  rise above the universal absorption $\pi e^2/2 h$ , starting in the visible spectral region.  There is a pronounced peak at $E=4.62eV$, dropping in the far UV region as in Figure \ref{absorption_fit}. Our calculations as described above and in he Supplementary Materials, with coordinate dependence of the transition moment included,  show that in the near-IR region, phononless direct absorption still dominates.  Starting in the visible, phonon-assisted absorption starts to play an increasingly important role, and is responsible for the quadratic rise in the near UV, with its influence increasing  into  the UV.  The  early rise starting in the visible   is not justified  by nonlinearities in the Dirac cones, but nonlinearities  do play a role at higher energies. 

Assuming linear electronic dispersion,  the density of electronic states for graphene is proportional to energy. For vertical phononless direct transitions, at a laser frequency $\hbar\omega$, an electron at $\hbar\omega/2$ below the Fermi level is excited to an empty state at $\hbar\omega/2$ above the Fermi level.   There is a lone  eligible conduction band electronic state for each occupied valence state, and the total number of vertical transitions is proportional to $\omega$.  

For phonon assisted indirect transitions, given a laser frequency $\hbar\omega$, the conduction band state can lie  anywhere with $0 < e<\hbar\omega$, and the hole can lie anywhere with energy $\hbar\omega-e > 0$ below the Fermi level, as long as the vertical energy gap between any given e-h pair is  $\hbar\omega$ minus the energy to create the associated phonon.    The total number of phonon assisted processes is proportional to $\int_0^{\hbar\omega}e(\hbar\omega-e)de \propto \omega^3$. Even though the matrix element for each indirect process is small compared to that of a direct one, the cubic growth of the number of indirect processes with energy makes the phonon assisted contribution significant at higher laser frequency. The two processes are shown in Figure \ref{absorption_emission_barePlus}, left. In the absorption calculation, there is a $\frac{1}{\omega}$ factor, making the contribution of direct  processes  constant   linear Dirac cone dispersion region.  The contribution of the phonon-assisted processes is seen rising at first  as $\omega^2$.  The nonlinearity of the Dirac cone in the UV region enlarges the electronic density of states,  contribute significantly to the UV excess absorption above 3 eV. 

The calculations are based on a supercell with periodic boundary conditions, giving a uniform sampling in the  Brillouin zone. There are sampling errors especially in the low energy region, where the number of states  is insufficient to compensate the $\frac{1}{\omega}$ factor accurately, as seen  in figure~\ref{absorption_fit} .  There is also a small error  compared to the known universal optical absorption  in the low energy region in Figure \ref{absorption_fit}, but the fit is good enough to give us confidence in the numerics.



\section{Ultrafast Hot Electronic Dynamics Upon Absorption}

\subsection{Phonon-Assisted Emission Spectrum}

It is found in several experiments that "relaxed" photoluminescence takes place in the time scale of 10 femtoseconds\cite{PhysRevB.82.081408,PhysRevLett.105.127404,PhysRevLett.102.086809}. These experiments were interpreted assuming both the absorption process and the emission process are purely phononless direct transitions. Thus the excited electron and hole distribution is extremely non-thermal. To fit the experimental results, the argument that ultrafast relaxation of electrons and holes, reaching a thermal distribution from an extremely non-thermal one in $5-10$ femtoseconds, is used. Unfortunately, the thermalization argument is compromised in that 2/3 of the electron energy has to be arbitrarily removed.

  The time scale for an electron-electron scattering process is in the order of several femtoseconds, and that for an electron-phonon scattering is in the order of picoseconds. Both time scales are too long to make the excited electrons and holes to reach full thermalization in the order of 10 femtoseconds.  To fit the experimental emission spectrum, a  temperature  much lower than the temperature of the excited electrons should reach  is used, otherwise there is a  ludicrous  9000 Kelvin  fit to the data.  This shedding of electronic energy   is not explained\cite{PhysRevLett.105.127404}. Under the thermalization argument, the   temperature should only depend on the incident laser fluence and not   the laser frequency. Instead  the data  shows  the  higher the frequency, the higher the temperature\cite{PhysRevB.82.081408}.

In our theory the "thermal" electron and hole distribution is produced nascently at $t = 0$ by a dominance of indirect transitions which, although present for weak radiation, take over from the easily bleached direct transitions in a bright pulse\cite{bao2009atomic}. No relaxation is required of the nascent distribution to give the observed emission spectrum. To state it plainly, the electron-hole distribution is "born" pre- "relaxed". Furthermore, the emission thereafter is almost certainly via vastly predominant inelastic indirect channels, which are always available from any electron to any hole, and do not need to wait perfect momentum coincidence of randomly distributed electrons and holes.

  The linear dispersion is assumed in our model. Assuming holes in the valence band and electrons in the conduction band are generated by phonon-assisted absorption and we ignore the matrix element for each transition for simplicity. Then the probability of a conduction band state located at $e$ above Fermi level occupied by an electron is proportional to the density of electronic states at $E_i-e$:
\begin{equation}ÒÒf(e)\propto(E_i-e)H(E_i-e)H(e)
\end{equation}
where $E_i$ is the incident light energy, and $H(x)$ is the Heaviside function. Similarly, the probability of a valence band state located at $e$ below Fermi level occupied by a hole is proportional to the density of states at $E_i-e$:
\begin{equation}
h(e)\propto(E_i-e)H(E_i-e)H(e)
\end{equation}

\begin{figure}[H]
\includegraphics[width=10cm]{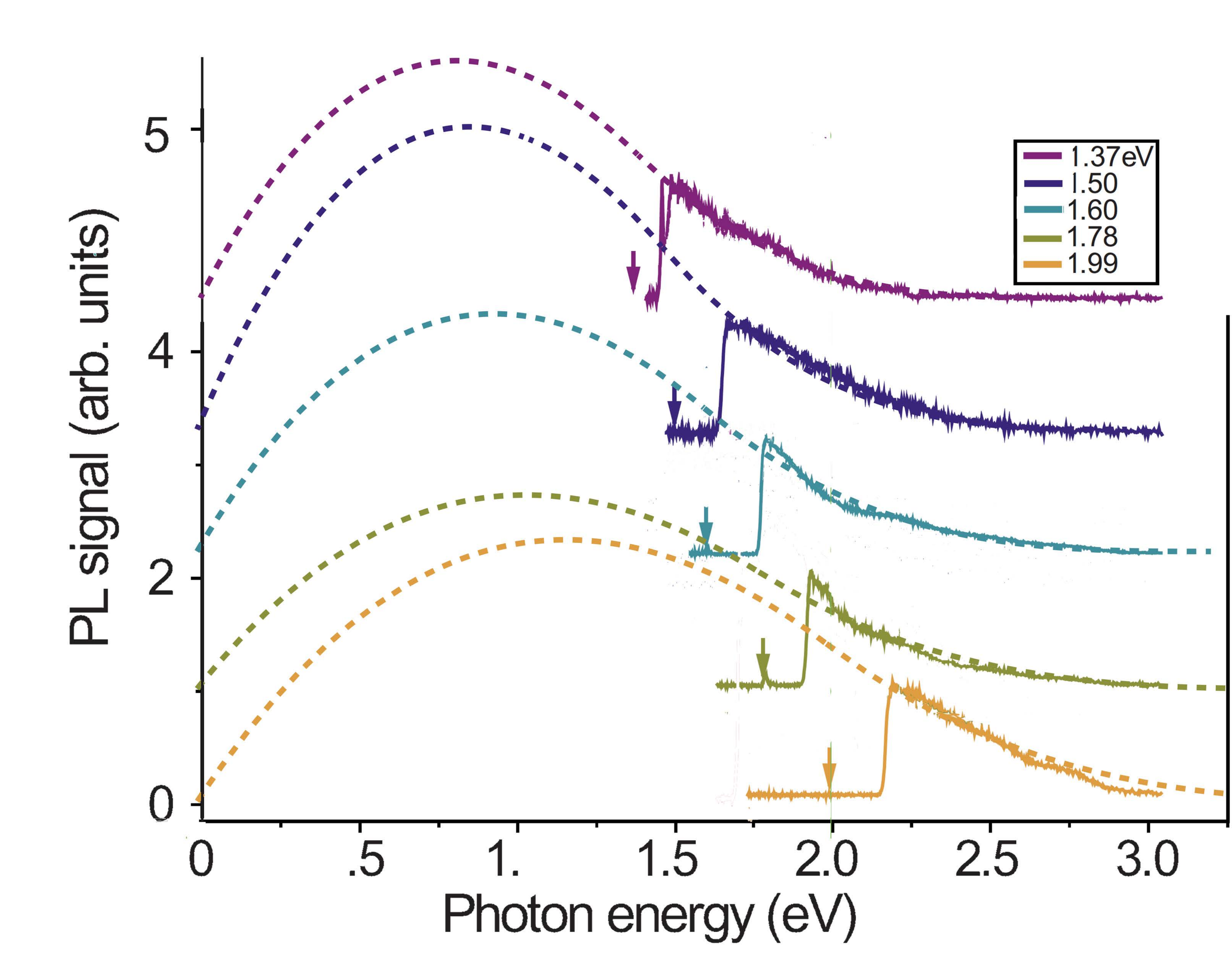}
\caption{Emission Spectrum with Different Incident Laser Frequencies: Solid lines are experiment results of ultrafast photoluminescence from graphene under different excitation photon energies adapted from Liu\cite{PhysRevB.82.081408}. Dashed lines are emission curves obtained from equation~\ref{emission_formula}. No relaxation of the the nascent electron-hole distribution or arbitrary energy removal is needed. Indirect absorption and emission, as in figure~\ref{absorption_emission_barePlus},  was used to calculate the emission spectrum, without adjustable parameters (save for the  vertical scale, which was arbitrary in the experiment). Note two important features not explained by the ultrafast thermalization model, but predicted by the indirect transition model: (1) The emission does not extend beyond energy $2 \hbar \omega$ (because that is the maximum indirect process separation of electrons and holes).  (2) There is movement of the high energy tail toward the UV as incident frequency is raised in the indirect mechanism, but the tail should be only fluence dependent, not frequency dependent, in the nearly instant thermalization model. }
\label{emission_fitwithphonon_process}
\end{figure}

Similarly, we ignore the matrix element variation for different transitions, simulating the emission by simple process counting. Then the phonon-assisted emission intensity   at $E_e$ when incident energy is $E_i$ by processes counting is:

\begin{equation}
\begin{split}
\sigma(E_e)&\propto\int_0^{E_i}\int_0^{E_i}e_1f(e_1)e_2h(e_2)\delta(E_e-(e_1+e_2))de_1de_2
\end{split}
\label{emission_formula}
\end{equation}
where the total number of excited electrons at $e$ above the fermi level is $\propto ef(e)$ and the total number of holes at $e$ below the fermi level is $\propto eh(e)$. We plot Eq (\ref{emission_formula}) for different $E_i's$, and a fit of the curve to experimental data is as Figure \ref{emission_fitwithphonon_process}. 







 
The expected  laser fluence $A$ dependence of the emission rate,  assuming a purely indirectly produced e-h pump population and indirect emission, goes as   $A^2$, exactly as seen in the experiments\cite{Mak20121341} at higher fluences. (Any electron able to emit to any hole with both populations proportional to $A$). The experiments show an $\sim A^{2.5}$ dependence at lower fluences.  This may be due to the onset of indirect process dominance as saturation becomes important.  The indirect processes are too numerous to get saturated, but if the source of the electrons and holes is a direct process, saturation would cut off the quadratic $A^2$ rise at larger pump fluences. This fact alone weighs heavily against the "direct transitions followed by ultrafast relaxation'' model. 

\subsection{Pump-probe spectra}

In pump-probe experiments by Breusing et. al., a starkly different probe  absorption spectrum from that expected from the presumed pump e-h population leads to the understandable conclusion that ultrafast relaxation must have taken place in the femtoseconds between pump and probe.  There is a totally different explanation involving indirect transitions  that fits the data extremely well.

\begin{figure}[htbp] 
   \centering
   \includegraphics[width=6.5in]{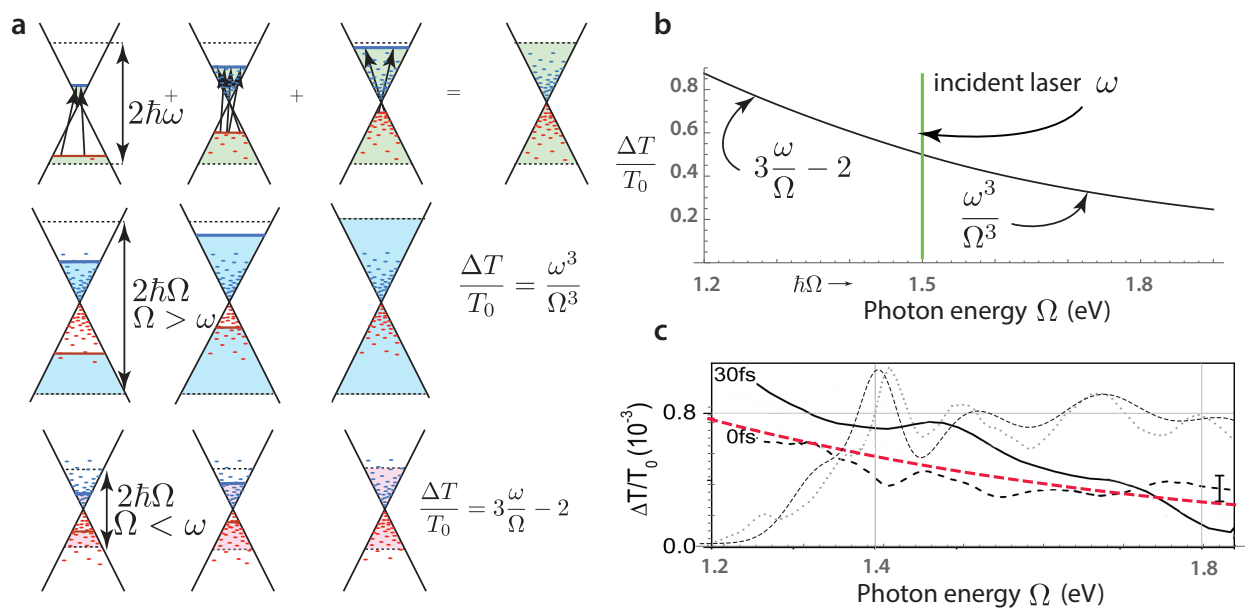} 
   \caption{ Explaining Breusing \textit{et.al}\cite{breusing2011ultrafast} results without the need for relaxation.  a) (Top row) Production of the nascent e-h distribution by indirect processes in the pump stage. The dashed lines give the upper  and lower  bounds  for the creation of electrons and holes at photon  energy $\hbar\omega$, respectively.  Transitions at $\hbar\omega$ (the total difference energy including  created or destroyed phonons) connect  the red and  blue lines as they slide over the Dirac cones.      The middle  row shows the range of  probe transitions at higher energy   $\hbar\Omega > \hbar\omega$; all the pump-produced electrons and holes separately induce missing processes in the probe pulse, inducing  $\Delta T/T \propto \omega^3/\Omega^3$.  In the bottom row, a lower energy probe photon  $\hbar\Omega < \hbar \omega$ picks up only some of the electrons and holes produced by the pump  as missing processes resulting in    $\Delta T/T \propto 3 \omega/\Omega - 2$. In b), the predicted probe absorption spectrum given a narrow band pump laser   at $\hbar \omega$ is shown as a black line; there is no vestige of the pump profile even with no relaxation after the pump c) a red dashed line shows the predicted probe spectrum assuming the first pulse had the spectral distribution of the thin dashed line; the experimental distribution is the grey dotted line. The experimental probe absorption spectrum is shown as dashed at 0 fs  and solid at  30 fs.  }
   \label{fig:pulsedThenPullsedAbsorption}
\end{figure}

The scenario is shown in figure~\ref{fig:pulsedThenPullsedAbsorption}. Even a   narrow bandwidth pump  gives a probe absorption spectrum looking very much like the broad and dispersed experimental one (panel b in the figure), without any relaxation required.   

There is no doubt that direct transitions are present,  and we now address their impact. Because we have questioned the existence of ultrafast relaxation of the 10 fs variety, we   assume the direct transitions are not much relaxed in the fastest experiments.  For  practical reasons, pump-spontaneous emission experiments must block emission at the incident laser wavelength, so they are not  revealing about any direct emission from unrelaxed direct absorption.  Pump-probe absorption is free of this problem, and indeed if  direct probe absorption followed unrelaxed direct  pump absorption then there should be an imprint of the pump profile on the probe. However if 20\% of the absorption is direct, and 20\% of the probe is also direct, then only 4\% of the probe absorption is  direct-direct.  The indirect  signature of a direct component is a very diffuse transparency  spectrum.  Of greatest importance in this discussion is that the femtosecond pump and pump-probe experiments appear to be run deep in the saturated absorption regime (see, for example, \cite{bao2009atomic}).

An earlier pump-probe experiment on graphite, not privy to the relaxed-appearing spectral distribution,  found a strongly bi-exponential decay, associating the faster decay with e-e relaxation,  stating ``the electronic system approaches an internal equilibrium with a characteristic time constant of  250 $\pm$ 50  fs''\cite{moos2001anisotropy}.  The slower ps and longer decay was associated with e-ph scattering.  In fact it needs stating that the more recent experiments also see these two timescales.  They are therefore stating  there are three relaxation times scales,  a sub 5-20 fs timescale, a 100-250 fs timescale, and a  1$+$ ps timescale. Here, we are maintaining the  sub 5-20 fs timescale does not exist, and is an erroneous but very understandable  inference from the deceptive indirect nascent e-h distribution. 

%

\section{Conclusion}
There are two major goals of this work.  The first  is to provide evidence that there is no ultrafast carrier relaxation in graphene implied by several important  experiments, in spite of appearances. This is   a matter of interpretation backed by the theory and calculations presented here, not a criticism of the experiments.  Second, we seek to   release the electronic transition moment from its traditional bondage (a principle that  also drove  re-casting the theory of Raman scattering in graphene\cite{heller2016theory}).  If the electronic transition moment depends on phonon displacements (as it must), then phonons are produced (or destroyed) the instant a photon is absorbed, within first order light matter perturbation theory.

Although the traditional frozen transition moment is badly misleading in graphene.  It is surely   a reasonable approximation in many other situations, but they  might profitably be re-examined with the electronic transition moment set free. A prime example are the indirect gapped transitions so important in many solids: it is known of course that a phonon is required to make these transitions allowed.  Nearly every source we have checked leaves the matter there, as if a necessity or a momentum conservation law is  an explanation of mechanism, which it is not. It is likely that  an explanation is  the  phonon coordinate dependence of the electronic transition moment.

\begin{acknowledgement}

This work was supported by the STC Center for Integrated Quantum Materials, NSF Grant No. DMR-1231319. The authors thank professor Efthimios Kaxiras, Wei Chen, Shiang Fang, Wenbo Fu, Ping Gao for helpful discussions.


\end{acknowledgement}

\section{Supplementary Material}

The formula for the total absorption cross section $\sigma$, for incident frequency $\omega$ and polarization $\hat{\boldsymbol{\epsilon}}$, between initial Born-Oppenheimer state $\ket{i}$ and final Born-Oppenheimer state $\ket{n}$ reads
\begin{equation}
\label{absorption_formula}
\sigma_{i,n}^{\hat{\boldsymbol{\epsilon}}}=\frac{4\pi^2\hbar}{m_e^2\omega}\frac{e^2}{\hbar c}\mid\bra{n}e^{i(\omega/c)(\hat{\boldsymbol{n}}\cdot\boldsymbol{x})}\hat{\boldsymbol{\epsilon}}\cdot\boldsymbol{p}\ket{i}\mid^2\delta(E_n-E_i-\hbar\omega)
\end{equation}

Making it more explicit, suppose with phonon coordinates $\boldsymbol{\xi}$ and electron coordinates $\boldsymbol{r}$, we write

\begin{equation}
\begin{split}
\ket{i}&=\ket{\phi(\boldsymbol{\xi};\boldsymbol{r})}\ket{\chi_{\boldsymbol{m}_i}(\boldsymbol{\xi})} \\
\ket{n}&=\ket{\phi_{\boldsymbol{q}_v\boldsymbol{q}_c}(\boldsymbol{\xi};\boldsymbol{r})}\ket{\chi_{\boldsymbol{m}_n}(\boldsymbol{\xi})}
\end{split}
\end{equation}

$\ket{\phi(\boldsymbol{\xi};\boldsymbol{r})}$ is the approximation to the Born-Oppenheimer electron ground state based on a Slater determinant of valence electron spin orbitals; $\ket{\phi_{\boldsymbol{q}_v\boldsymbol{q}_c}(\boldsymbol{\xi};\boldsymbol{r})}$ is an electron-hole pair relative to the ground state, with an electron in the conduction band orbital $\ket{\phi_{\boldsymbol{q}_c}(\boldsymbol{\xi};\boldsymbol{r})}$ with momentum $\hbar \boldsymbol{q}_c$  and a hole in the valence band orbital $\ket{\phi_{\boldsymbol{q}_v}(\boldsymbol{\xi};\boldsymbol{r})}$ with momentum $\hbar \boldsymbol{q}_v$.  $\ket{\chi_{\boldsymbol{m}_i}(\boldsymbol{\xi})}$ and $\ket{\chi_{\boldsymbol{m}_n}(\boldsymbol{\xi})}$ are phonon wavefunctions in the initial state and final state respectively. In the case of a lattice, each phonon coordinate is independent of each other. The potential of each phonon could be well approximated by a harmonic potential, and the wavefunction for each phonon coordinate takes the form of a gaussian multiplied by an Hermite polynomial. $\boldsymbol{m}_n$ and $\boldsymbol{m}_i$ are series of quantum numbers indicating which phonon is excited. In the initial state, we treat phonons are in ground states, then $\boldsymbol{m}_i=(0,0,...,0)$. Under electric dipole approximation and with the Born-Oppenheimer approximation, the matrix element in Eq (\ref{absorption_formula}) can be expressed as 

\begin{equation}
\begin{split}
&\bra{n}e^{i(\omega/c)(\hat{\boldsymbol{n}}\cdot\boldsymbol{x})}\hat{\boldsymbol{\epsilon}}\cdot\boldsymbol{p}\ket{i}\\
\approx&\bra{n}\boldsymbol{D}^{\hat{\boldsymbol{\epsilon}}}\ket{i}=\bra{\chi_{\boldsymbol{m}_n}(\boldsymbol{\xi})}\bra{\phi_{\boldsymbol{q}_v\boldsymbol{q}_c}(\boldsymbol{\xi};\boldsymbol{r})}\boldsymbol{D}^{\hat{\boldsymbol{\epsilon}}}\ket{\phi(\boldsymbol{\xi};\boldsymbol{r})}\ket{\chi_{\boldsymbol{m}_i}(\boldsymbol{\xi})}
\end{split}
\end{equation}

Rewrite $\bra{n}\boldsymbol{D}^{\hat{\boldsymbol{\epsilon}}}\ket{i}$ as

\begin{equation}
\bra{n}\boldsymbol{D}^{\hat{\boldsymbol{\epsilon}}}\ket{i}=\bra{\chi_{\boldsymbol{m}_n}(\boldsymbol{\xi})}\mu_{\boldsymbol{q}_c,\boldsymbol{q}_v}^{\hat{\boldsymbol{\epsilon}}}(\boldsymbol{\xi})\ket{\chi_{\boldsymbol{m}_i}(\boldsymbol{\xi})}
\end{equation}
with
\begin{equation}
\mu_{\boldsymbol{q}_c,\boldsymbol{q}_v}^{\hat{\boldsymbol{\epsilon}}}(\boldsymbol{\xi})=\bra{\phi_{\boldsymbol{q}_v\boldsymbol{q}_c}(\boldsymbol{\xi};\boldsymbol{r})}\boldsymbol{D}^{\hat{\boldsymbol{\epsilon}}}\ket{\phi(\boldsymbol{\xi};\boldsymbol{r})}_{\boldsymbol{r}}
\end{equation}

The matrix elements of the dipole operator $\boldsymbol{D}^{\hat{\boldsymbol{\epsilon}}}$ between two Born-Oppenheimer electronic states is the transition moment $\mu_{\boldsymbol{q}_c,\boldsymbol{q}_v}^{\hat{\boldsymbol{\epsilon}}}(\boldsymbol{\xi})$ connecting valence level $\boldsymbol{q}_v$ and conduction band electronic levels $\boldsymbol{q}_c$. The subscript $\boldsymbol{r}$ indicates that only the electron coordinates are integrated. Note that $\mu_{\boldsymbol{q}_c,\boldsymbol{q}_v}^{\hat{\boldsymbol{\epsilon}}}(\boldsymbol{\xi})$ is a function of phonon coordinates $\boldsymbol{\xi}$. If we perform Taylor expansion of the transition moment around the equilibrium geometry of the lattice till the first order term of lattice distortion, we have

\begin{equation}
\label{TM_Tayor_Expansion}
\mu_{\boldsymbol{q}_c,\boldsymbol{q}_v}^{\hat{\boldsymbol{\epsilon}}}(\boldsymbol{\xi})=\mu_{\boldsymbol{q}_c,\boldsymbol{q}_v}^{\hat{\boldsymbol{\epsilon}}}(\boldsymbol{\xi}_0)+\sum_j{\frac{\partial{\mu_{\boldsymbol{q}_c,\boldsymbol{q}_v}^{\hat{\boldsymbol{\epsilon}}}(\boldsymbol{\xi})}}{\partial \xi_j}}\mid_{\boldsymbol{\xi}=\boldsymbol{\xi}_0}\mid\boldsymbol{\xi}_j-\boldsymbol{\xi}_0\mid+O(\mid\boldsymbol{\xi}_j-\boldsymbol{\xi}_0\mid^2)
\end{equation}

\subsection{Vertical Phononless Absorption (Direct Processes)}
We first consider  only the constant part  of the transition moment $\mu_{\boldsymbol{q}_c,\boldsymbol{q}_v}^{\hat{\boldsymbol{\epsilon}}}(\boldsymbol{\xi}_0)$   Eq (\ref{TM_Tayor_Expansion}), i.e.

\begin{equation}
\label{Direct_Absorption}
\sigma_{i,n}^{\hat{\boldsymbol{\epsilon}},direct}=\frac{4\pi^2\hbar}{m_e^2\omega}\frac{e^2}{\hbar c}\mid\bra{\chi_{\boldsymbol{m}_n}(\boldsymbol{\xi})}\mu_{\boldsymbol{q}_c,\boldsymbol{q}_v}^{\hat{\boldsymbol{\epsilon}}}(\boldsymbol{\xi}_0)\ket{\chi_{\boldsymbol{m}_i}(\boldsymbol{\xi})}\mid^2\delta(E_n-E_i-\hbar\omega)
\end{equation}

Promoting a small minority of the valence electrons to the conduction band leaves the lattice constants almost unchanged.  The vast majority of terms in Eq (\ref{Direct_Absorption}), vanish, due to zero phonon wavefunction overlap and momentum mismatch. Of all the possible excited states,  only a few will have the right energy and fewer still have the right momentum. So we can further simplify Eq (\ref{Direct_Absorption}). 

For simplicity we suppose that the lattice is initially in the phononless ground state, namely $\chi_{\boldsymbol{m}_i}(\boldsymbol{\xi})=\chi_{(0,0,...,0)}(\boldsymbol{\xi})$. Since there are no geometry changes involved in the process (constant transition moment), there would be no phonons generated in the final state, and the transition is purely electronic transition. Since there are no phonons involved, and the momentum of light is ignored under dipole approximation, the transition between electronic state must conserve momentum. So in Eq (\ref{Direct_Absorption}), we have $\chi_{\boldsymbol{m}_n}(\boldsymbol{\xi})=\chi_{\boldsymbol{m}_i}(\boldsymbol{\xi})=\chi_{(0,0,...,0)}(\boldsymbol{\xi})$ and $\boldsymbol{q}_c=\boldsymbol{q}_v$, which gives vertical phononless transition, namely
\begin{equation}
\sigma_{i,n}^{\hat{\boldsymbol{\epsilon}},direct}=\frac{4\pi^2\hbar}{m_e^2\omega}\frac{e^2}{\hbar c}\mid\bra{\chi_{(0,0,...,0)}(\boldsymbol{\xi})}\mu_{\boldsymbol{q}_c,\boldsymbol{q}_v}^{\hat{\boldsymbol{\epsilon}}}(\boldsymbol{\xi}_0)\ket{\chi_{(0,0,...,0)}(\boldsymbol{\xi})}\mid^2\delta(\boldsymbol{q}_c-\boldsymbol{q}_v)\delta(E_n-E_i-\hbar\omega)
\end{equation}

So we can reduce the summing over all possible states (including phonon excitation) to summing over states with electron momentum and energy matched, and phonon in the ground state to get the total vertical phononless absorption. Similar arguments can also be applied to emission formula.

\subsection{Phonon-Assisted Absorption (Indirect Processes)}
We consider the first order terms in Eq (\ref{TM_Tayor_Expansion}), and these terms give phonon-assisted transition. Similarly we assume that there is no phonon in the initial state, namely $\chi_{\boldsymbol{m}_i}(\boldsymbol{\xi})=\chi_{(0,0,...,0)}(\boldsymbol{\xi})$. The first order term of the transition moment implies the dependence of the transition moment on the $jth$ phonon coordinate, which means a type $j$ phonon could be generated in the process. So in the final state we have phonon wavefunction $\bra{\chi_{(0,0,...,1,...,0)}(\boldsymbol{\xi})}$,where (0,0,...,1,...,0) means the $jth$ phonon is excited to the first excited state. The transition to higher vibrational states could be ignored because their transition matrix elements are much smaller. So we have \begin{equation}
\sigma_{i,n}^{\hat{\boldsymbol{\epsilon}},indirect}=\frac{4\pi^2\hbar}{m_e^2\omega}\frac{e^2}{\hbar c}\mid\bra{\chi_{(0,0,...,1,...,0)}(\boldsymbol{\xi})}\frac{\partial{\mu_{\boldsymbol{q}_c,\boldsymbol{q}_v}^{\hat{\boldsymbol{\epsilon}}}(\boldsymbol{\xi})}}{\partial \xi_j}\mid_{\boldsymbol{\xi}=\boldsymbol{\xi}_0}(\xi_j-\xi_{j,0})\ket{\chi_{(0,0,...,0)}(\boldsymbol{\xi})}\mid^2\delta(E_n-E_i-\hbar\omega)
\end{equation}
Let $\boldsymbol{k}_j$ be the wave vector of the $jth$ phonon. So under electric dipole approximation, we need $\boldsymbol{q}_c=\boldsymbol{q}_v+\boldsymbol{k}_j$ to conserve momentum, namely

\begin{equation}
\begin{split}
\sigma_{i,n}^{\hat{\boldsymbol{\epsilon}},indirect}=&\frac{4\pi^2\hbar}{m_e^2\omega}\frac{e^2}{\hbar c}\mid\bra{\chi_{(0,0,...,1,...,0)}(\boldsymbol{\xi})}\frac{\partial{\mu_{\boldsymbol{q}_c,\boldsymbol{q}_v}^{\hat{\boldsymbol{\epsilon}}}(\boldsymbol{\xi})}}{\partial \xi_j}\mid_{\boldsymbol{\xi}=\boldsymbol{\xi}_0}(\xi_j-\xi_{j,0})\ket{\chi_{(0,0,...,0)}(\boldsymbol{\xi})}\mid^2\\
&\delta(E_n-E_i-\hbar\omega)\delta(\boldsymbol{q}_c-\boldsymbol{q}_v-\boldsymbol{k}_j)
\end{split}
\end{equation}
The sum  over all possible states is reduced to those  states with one phonon excited, with electronic to states making up for the  momentum  and energy of the phonon, leaving energy and momentum conserved. Similar arguments can also be applied to emission formula.

In support of the importance of phonon-assisted processes, reference~\cite{nonSatAbs} reports a nonsaturable absorption component in graphene of about 34\% for 2-4 layers of graphene.  The indirect transitions are extremely difficult to saturate, given their huge variety and number, and very likely the cause of the nonsaturable component, there being no other transitions in this relevant frequency range,

\bibliography{achemso-demo}

\end{document}